\documentclass[conference,letterpaper]{IEEEtran}

\usepackage[top=0.75in, bottom=1.05in, left=0.75in, right=0.75in, heightrounded]{geometry}
\usepackage{setspace}
\usepackage{amsmath} 

\usepackage[hyphens]{url}
\usepackage{hyperref}
\hypersetup{colorlinks,allcolors=blue}

\usepackage[backend=bibtex,style=ieee,citestyle=ieee-comp]{biblatex}
\usepackage{comment}
\IEEEoverridecommandlockouts

\usepackage{graphicx} 
\usepackage{xcolor} 
\usepackage{minted} 
\usepackage{doi}
\usepackage{orcidlink}
\def\BibTeX{{\rm B\kern-.05em{\sc i\kern-.025em b}\kern-.08em
    T\kern-.1667em\lower.7ex\hbox{E}\kern-.125emX}}

\begin{document}

\title{Confetti Magic: A Case Study of a China-Nexus Operational Relay Box (ORB) Network
\thanks{Disclaimer: The views expressed are those of the authors and do not necessarily reflect the official policy or position of the U.S. Department of Defense or the U.S. Government. References to external sites do not constitute endorsement. Cleared for release on 22 OCT 2026 (DOPSR 27-T-0171).}
}

\author{\IEEEauthorblockN{Adam Dorian Wong\,\orcidlink{0009-0000-1832-6859}, Cody Welu}
\IEEEauthorblockA{\textit{The Beacom College of Computer \& Cyber Sciences} \\
\textit{Dakota State University}\\
Madison, SD, USA \\
adam.wong@trojans.dsu.edu, cody.welu@dsu.edu
}
}

\maketitle

\begin{abstract} 
Advanced Persistent Threat (APT) groups operating out of the People's Republic of China (PRC) are increasingly reliant on  Operational Relay Box (ORB) proxy networks to obfuscate their origins and evade detection. Despite these recent developments, ORB networks are largely under-researched in academic literature. This paper presents a qualitative observational case study, \textit{"Confetti Magic"}, which analyzes the technical capabilities and infrastructure patterns of APT Group UAT-7810, a known ORB provider tracked by Cisco Talos. By tracking artifact reuse, examining behavioral patterns, and utilizing open-source intelligence tools, this study identifies a newly reconstituted ORB hop-point. Based on documentary evidence, observations, and geographic evidence, APT Group UAT-7810 is likely a private contractor associated with the Jiangsu State Security Department (SSD) of the Chinese Ministry of State Security (MSS). This research provides an example of identifying ORB hop-points based on threat actor behaviors and brittle artifact reuse.
\end{abstract}


\section{Introduction}
Adversary tactics, techniques, and procedures (TTPs) consistently evolve to meet the requirements of a changing digital battle space. Defenders rely upon standard artifacts like logs, flow records, and packet captures (PCAP) to understand the scope of an intrusion. Advanced Persistent Threat (APT) groups actively utilize anti-forensic countermeasures to obfuscate attack origins and evade Cyber Threat Intelligence (CTI) analyses. A recent development is the application of complex, multi-layer proxy networks composed of compromised Internet-of-Things (IoT) botnets, procured Virtual Private Servers (VPS), or commercial Virtual Private Network (VPN) services.

The People's Republic of China has cultivated its cyber operations talent via a complex ecosystem and hacker culture among the People's Liberation Army (PLA), Ministry of State Security (MSS), Ministry of Public Security (MPS), academia, contractors, and private citizens. These operations are perceived as congruent with their Five-Year Plans. In 2014, then-FBI Director James Comey compared the average Chinese APT to a \textit{"drunken burglar"} \cite{thomas_comey_2014}. A decade later, the Chinese cyber operation \textit{modus operandi} has prioritized stealth in long-term engagements.

APT groups aim to prevent observation and reduce traceability through the use of botnets and proxy networks to relay network traffic. This technique has been dubbed the \textit{Operational Relay Box (ORB)} network. ORB networks are becoming common among Chinese cyber operations. ORB networks are poorly understood and are largely unaddressed in academic literature.

To address this gap, this qualitative observational case study examines a set of ORB hop-points constructed, terminated, and reconstituted by Chinese APT UAT-7810 (a Cisco moniker) in 2026. The ORB hop-points studied are referred by Cisco Talos as the "LapDogs" ORB network. By analyzing this infrastructure provider's technical capabilities, this research aims to link the real-world identities to monikers, providing valuable insight into both the malicious infrastructure provider UAT-7810 (Cisco) as well as the offensive cyber operators they support, tracked as APT Group UAT-5918 (Cisco).

This paper makes three primary contributions:
\begin{enumerate}
    \item Provides deeper insight into an example of ORB hop-points through monitoring of open directories, during an active campaign.
    \item Attribution of several new hop-points to UAT-7810 through artifact and IP analyses.
    \item Suggests evidence that UAT-7810 (LapDogs ORB provider) is a contractor for the Jiangsu State Security Department (SSD) of the Chinese Ministry of State Security (MSS) or People's Liberation Army Cyberspace Force (PLA-CSF) Eastern Theater Command.
\end{enumerate}

The remainder of this paper frames the case study through the literature review \& research gap, methodology, findings, and discussion of implications, recommendations, future work, and limitations.

\section{Related Work}
\subsection{Academic Literature}
Previous efforts in researching APT groups vary greatly. Researchers have attempted to understand APT campaigns and defend against them. In 2025, Chen's team examined use of deep reinforcement learning (RL) to defend a network \cite{chen_defending_2025}. However, this study utilized a small testbed environment with limited realism and generalization. Zuo conducted a survey of complex intrusions and re-evaluated the APT lifecycle \cite{zuo_empirical_2025}. This study lacked realism as well.
 
Proper attribution is a key concern. However, current research has incorporated machine learning (ML) techniques to enhance security analyst experiences. Hwang conducted a qualitative study in 2023 examining aspects of attribution and false flag operations \cite{hwang_exploratory_2023}. Now, Hwang's use of white-hat hackers does not necessarily reflect the real-world conditions with APT operators with lower ethical barriers. Lee and Kim utilized generative neural networks to sequence TTPs with intent to create intrusion kill chains of varying style and build a larger dataset \cite{lee_mucamp_2025, kim_mucamp2_2026}. It lacked transferability because different groups use different techniques and the primary dataset was the Lazarus Group. Ge's research team created ThreatMAMBA to help assist in attribution based on partial kill chains \cite{ge_threatmamba_2026}.

Covert communications is another theme congruent with APT behaviors. Zou's research in covert channels differs where compromised networking hardware may be used as a proxy for command-and-control (C2) \cite{zou_passive_2026}. One limitation of Zou's simulation is that it lacked use of encryption which is common in intrusions by commodity malware and APTs alike.

There is some research in covert communications, but it is in radio-layer theory. Su's research presented a mathematical model in a wireless network where relays send covert messages \cite{su_covert_2021}. This research is very limited in relation to cyber operations. Ding also examined radio-layer theory, but one key finding was that threat actors may only act when their activities blend with normal traffic \cite{ding_covert_2025}. Similar to Su, this has very limited transferability to APT intrusions.

Research has also been conducted around botnet infrastructure. Shashwat incorporated large language models (LLMs) into similarity searches in TLS certificates with the assumption that attackers lazily generate certificate fields \cite{10.1145/3689932.3694766}. One limitation is that the threat detection relies solely on TLS certificate fields as an indicator of nefarious activity. Alomari takes a different approach by creating a detector driven by flow records and augmented with explainable AI (XAI)  \cite{10.1145/3727967.3756837}. Similarly, the internal validity is dependent on ports and flow records.

In 2015, Poornachandran examined domain name system (DNS) hijacking with small office home office (SOHO) and Internet-of-Things (IoT) devices and pointed to hardware life cycle as a contributing factor to cyber attacks \cite{poornachandran_internet_2015}. The internal validity may be questioned since it conflates suspicious with malicious. Zhang evaluated Tor exit nodes through server-side analyses and concluded that blocks could be circumvented by exit bridges \cite{zhang_ephemeral_2020}. Jain assessed Internet Service Provider (ISP) furnished consumer-grade SOHO routers and found these devices are poorly configured or even completely unsecured \cite{jain_accounting_2023}.

\subsection{Industry Literature}
Outside of academia, industry is focused on threat intelligence with some reporting around ORB networks. Digital forensics has evolved since IETF RFC3227 \cite{brezinski_guidelines_2002}. In 2012, Jason Healey defined a \textit{"Spectrum of State Responsibility"} to define the ownership of actions by a nation and perceived threat actors \cite{healey_beyond_2012}. David Bianco proposed the \textit{"Pyramid of Pain"} to give weighted value to indicators-of-compromise (IOCs) \cite{bianco_enterprise_2013}. Data leaks have been used to analyze Chinese contractors (e.g.: i-S00n) working on behalf of the Chinese government \cite{cary_unmasking_2024}. Additionally, the anonymous bloggers of IntrusionTruth have consistently led attribution efforts for Chinese APTs \cite{noauthor_intrusion_2026}.

The PLA's National Defense University (NDU) textbook around strategic thought explicitly called for Military-Civil fusion or integration \cite{xiao_science_2022}. A Mandiant Threat Intelligence Report around Chinese cyber-espionage suggested that the PLA and MSS underwent reorganizations and have steadily improved their cyber capabilities since 2016 \cite{vanderlee_chinas_2022}. In 2023, Mandiant, now Google, continued to track Chinese cyber espionage operators increasingly relying on living-off-the-land (LoTL) techniques and botnets for traffic obfuscation \cite{mandiant_intelligence_stealth_2023}. This is one of the earliest observations and general acknowledgments in industry for botnets-as-a-proxy. A public Joint Cyber Advisory from the US Government (USG) warned about PRC hackers infecting routers and IoT devices to incorporate into their botnets \cite{federal_bureau_of_investigation_peoples_2024}. Raggi expanded this idea further and coined these botnets: "ORB networks" \cite{raggi_ioc_2024}.

Team Cymru began hunting for ORB networks following Raggi's announcement and suggested hunting ORB hop-points using certificate fingerprints or non-traditional open ports \cite{team_cymru_introduction_2024}. In 2025, a company called SecurityScorecard identified a specific ORB network dubbed: \textit{"LapDogs"} by pivoting on TLS fingerprints \cite{securityscorecard_checking_2025}. Thus, Cisco Talos further investigated and released several IP address indicators in July 2026. Cisco Talos tracked the suspected infrastructure provider as APT Group "UAT-7810". Industry vendors suspect that the LapDogs ORB network is likely used by operationally-focused APT groups like APT Group "UAT-5918" \cite{an_uat-7810_2026}. The appearance of one ORB network is not a one-off coincidence. This is a doctrinal evolution by the Chinese which enables their cyber attacks. ORB networks have also been seen by other attack groups such as Flax Typhoon \cite{noauthor_matter_2024, united_states_cyber__infrastructure_security_agency_defending_2026}.

\section{Research Gap} Based on available literature, it is clear that research varies across different themes such as: geopolitics, threat attribution, threat hunting, detections, covert channels, botnet infrastructure, and SOHO / IoT device security. Naturally, there is a gravitation on the incorporation of AI and ML models across these themes in academia. In industry, there is a divergence and a natural focus on threat hunting.
\begin{enumerate}
    \item There is very little discussion in academia around ORB networks which will be a persistent issue.
    \item There is limited literature tying nuances of these ORB hop-points to geo-political and social issues.
\end{enumerate}

The overall research objective is to analyze the technical capabilities, operational infrastructure, and behavioral artifacts of a newly discovered ORB network hop-point, constructed by UAT-7810, in an effort to establish a definitive link between the infrastructure provider (UAT-7810) and its real-world state-sponsored entity.

\section{Methodology}
\subsection{Research Design} This research follows a descriptive case-study around one particular ORB network newly discovered and undisclosed following review of the Cisco Talos report of July 2026. It utilizes documentary evidence collected from across various public and commercial cybersecurity tools or datasets. This qualitative case study was named: Confetti Magic. It follows purposive sampling to extract information and artifacts from suspicious IP addresses through commercially available tools.

\subsection{Case Bounding}
\begin{enumerate}
    \item This case study spans between November 2025 and October 2026.
    \item Four IP addresses were reported by Cisco Talos in July 2026 (Hop-Points 1-4). These four IP addresses generally appear active between late 2025 until approximately September 2026.
    \item As part of the Confetti Magic study, this newly discovered ORB IP address or ORB hop-point was first logged by Hunt.io in late September 2026.
\end{enumerate}

\subsection{Data Analysis}
\textbf{Discovery.} Between July 2026 and September 2026, the team noted that suspicious server ports previously open were now closed. In late September 2026, the research team started reviewing open directories of IP addresses (Hop-Points 1-4) reported by Cisco Talos and preserving their snapshots of contents from Hunt.io. Open directories contain publicly accessible files which the threat actors needed readily available. 

The team iterated and reviewed the contents of the new server and found that the naming conventions used across all the files were similar, even if the payloads varied from the originating servers.

At first glance, pivoting in Hunt.io's open directories by filenames is a very brittle indicator. Filenames are easily replaceable or internal file contents can be different. Threat actors are human beings and it is difficult to change behavior. David Bianco emphasized this in his \textit{Pyramid of Pain} in which burning TTPs imposes the greatest cost to threat actors in terms of investment of time and resources to reinvent \cite{bianco_enterprise_2013}.

The team time-boxed the Hunt.io search between July 2026 and October 2026 using the same naming schema. The team approached the search with the assumption that threat actors are not likely to change their filenames or TTPs until discovered by the public. Only one new IP address (Hop-Point 5) matched the exact filename seen on Hop-Point 4. Now, Hop-Point 5 was recorded by Hunt.io within a month following perceived sanitation of servers outlined in the Cisco Talos report, but it was unlabeled by Hunt.io. Hop-Points 1-4 closed their C2 ports throughout summer of 2026.

Hop-Points 6 and 7 were discovered after conducting open directory text searches and finding hard-coded references to the Hop-Point 5 IP addresses in text files.

The team iterated over each IP old (Cisco reported IOC) and new (discovered IOC) and archived IP timelines, WHOIS records, Shodan scans, files, and directory listings. The team also analyzed scripts from the directory listing to understand the scripts' purpose and behavior.

For geographic analysis, the team examined Chinese law summaries, open-source information around Chinese businesses, and map locations to perceive possible connections to state enterprises.

\subsection{Data Trustworthiness}
This information was corroborated through several tools such as manual code review for downloader payloads, DomainTools IRIS Investigate for WHOIS records, Shodan for port history records, Deception.Pro \& VirusTotal for artifact analyses, third-party business registration records, and open-source mapping software.

Both Hunt.io and Shodan overlapped around observed port values and approximate timelines which these IPs were active. These hop-points are hosted by different providers, but demonstrate a consistent theme of hosting web servers on unusual ports, script-based exploits, and multiple C2 proxy configuration files.

WHOIS and ASN records were validated between Hunt.io, Shodan, and DomainTools. The discovered hop-points were also checked against a public gray-market list of known proxy IP addresses operating in China. No matches were identified against the gray-market list.

From a data collection standpoint, hop-point IP addresses were processed identically and linearly from Hunt.io, DomainTools, and Shodan. Data was archived through both Zotero and page print-outs.

\subsection{Ethical Considerations} At the time of discovery, the research team validated the server's likely purpose after reviewing captured artifacts. Research was paused due to the likelihood that Hop-Points 5-10 posed an imminent danger to civilian infrastructure. Information was deferred to law enforcement officials via the FBI IC3 criminal complaint system.

\section{Findings}
\begin{table*}[t]
\centering
\caption{Hop-Points Analyzed}
\label{tab:hop_points}
\begin{tabular}{|c|c|c|c|}
\hline
Hop-Points & IP & ASN & Data Source \\ \hline
HP1 & 95.182[.]x.x & AS56971 (Cloud) & Cisco Talos \\ \hline
HP2 & 194.233[.]x.x & AS141995 (Contabo Asia) & Cisco Talos \\ \hline
HP3 & 217.15[.]x.x & AS141995 (Contabo Asia) & Cisco Talos \\ \hline
HP4 & 217.15[.]x.x & AS141995 (Contabo Asia) & Cisco Talos \\ \hline
HP5 & 103.43[.]x.x & AS132883 (Jiangsu Sanai NST) & Hunt.io  \\ \hline
HP6 & 103.43[.]x.x & AS132883 (Jiangsu Sanai NST) & Hunt.io \\ \hline
HP7 & 37.123[.]x.x & AS60024 (Kirino LLC) & Hunt.io  \\ \hline
HP8 & 114.255[.]x.x & AS4808 (China Unicom) & Hunt.io \\ \hline
HP9 & 51.75[.]x.x & AS16276 (OVH GmbH) & Hunt.io \\ \hline
HP10 & 8.218[.]x.x & AS45102 (Alibaba Cloud) & Hunt.io \\ \hline
\end{tabular}
\end{table*}

\textbf{Running Services.} Although ports were mostly different across Hop-Points 1-10, IPs were observed using non-traditional or non-standard alternative ports. Through banner grabbing via Hunt.io and Shodan, numerous HTTP web services ran on multiple obscure ports. A system may consist of outdated versions of nginx 1.24.0, SimpleHTTP/0.6 Python 3.10.12 or Python/3.8.16, or Apache web.

Team Cymru, SecurityScorecard, and Cisco Talos reported unusual certificate common names such as "exploit" \cite{team_cymru_introduction_2024, securityscorecard_checking_2025, an_uat-7810_2026}. The Confetti Magic study identified an unusual common name \textit{"win-https-8443"}.

For example, if a system has Ports 80 / 443 (HTTP/S) open, then it would also host crafted TLS certificates and run web services on unusual alternative ports such as Port 93, 99, 2222, 3000, 8088, 9999, 18081, and 18095 instead of common conventional alternate numbers like 8080 or 8000.

The new hop-points unveiled by this study appear to be tailored with a central theme or goal.
\begin{itemize}
    \item Hop-Point 5 is a Chisel C2 server.
    \item Hop-Point 6 is a Chisel Client with a theme of Windows scripts and executables. It also carries an AdaptixC2 payload as "packed.exe".
    \item Hop-Point 7 is a Clash C2 proxy server with SSDP scripts and target IP lists.
    \item Hop-Point 8 appears to be a staged system, not yet configured.
    \item Hop-Point 9 is used as a Windows error log staging area for Hop-Point 7.
    \item Hop-Point 10 is used as a destination for enumeration data, but packets from Hop-Point 7 are crafted with Scapy to spoof sender origins.
\end{itemize}

\textbf{Proxy Chains} The IP address for Hop-Point 5 was referenced in an obscure Virtual Network Computing (VNC) text file on Hop-Point 6. 

Hop-Point 6 contained numerous batch scripts, PowerShell scripts, and PowerSploit. It was a Chisel client for Hop-Point 5 through a PowerShell script called "wdc.ps1". This PowerShell script was configured to maintain a persistent Client connection to Hop-Point 5 (Chisel Server). Additionally, it was capable of establishing reverse tunnels within multiple private IP addresses and target services like HTTP and MySQL. A Base64 encoded file called "dv\_tok.txt" contained a JSON Web Token with dummy information, except the phone number which pointed to Tianjin BFS Building Material Tech. Co. Ltd. Hop-Point 6 referenced Hop-Point 5 as the destination contained in numerous beacon and tunnel scripts. Hop-Point 6 IP was also contained inside a file called "7001.txt" on Hop-Point 7. Additionally, Hop-Point 6's file "t22c.sh" contained multiple reverse tunnels to target intranet systems via Chisel.

Hop-Point 7 contained executables for Clash C2 proxy. It referred directly to Hop-Point 8 in a file of open VNC nodes. The open directory included several YAML configuration files for Clash tunnel proxies, hard-coded DNS configuration for Quad9.net, DNS-over-HTTP (DoH), and Simple Service Discovery Protocol (SSDP) amplification flooder tools. This IP address belonged to AS60024 or Kirino LLC. Now, some configuration files inside Hop-Point 7 held references to unusual generated (DGA) subdomains belonging to route AS37963 or Wuhan Piliang Management Co. Ltd.

Hop-Point 8 belonged to AS4808 (Beijing) under China Unicom. At the time of research, this hop-point had not yet been fully staged or configured. The team did not observe any open directories on Hop-Point 9-10.

The cyclical connections represent a concerted effort to obfuscate network traffic.

\textbf{Malicious Files} Several pieces of shell code and malware contained Chinese characters in comments. The files on Hop-Point 5 were highly suspicious due to containing renamed binaries for commodity malware or tools (e.g.: Mimikatz, JuicyPotato, CobaltStrike, and Meterpreter), LoTL firewall manipulation scripts, scanning scripts written in Python, VPN configurations, FTP credentials, SSH private keys, and SMB exploitation scripts (e.g.: MS17-010). This new ORB system was deployed with multiple exploits for use. Shell scripts on this hop-point also utilize common Bash commands for Linux targets. This represents a threat actor utilizing mis-attribution techniques.

Hop-Point 7 contained a Python script called "attack.py" and "attackv2.py". These scripts sent error logs to a Windows VPS host in Germany, while crafting packets with Scapy to send toward an Alibaba VPS IP as the destination. The attack script collects target hardware specifications, network speed test data, user or process permissions, and converts timestamps to the Beijing time zone.

\textbf{Network Ownership} Alone, the indicator is brittle. However, the team observed that among the Cisco indicators: 1x IP address belonged to route AS56971 CGI Global Ltd. (Hong Kong) and 3x IP addresses belonged to route AS141995 or Contabo Asia Private Ltd. (Singapore). Both of these Autonomous System Numbers (ASNs) belong to commercial hosting or leasing providers consistent with open-source reporting \cite{raggi_ioc_2024}.

Although Singapore is sovereign and Hong Kong is viewed as a semi-autonomous region, these areas are culturally significant and deeply intertwined with Chinese commerce \cite{ying_as_2025}. China's own series of national security and intelligence laws influence Chinese corporations, even when operating abroad \cite{cna_chinas_2023}. China tightly controls outbound network traffic through several metropolitan cities acting as international gateways with peer-nations, its national interconnect is mostly managed through 3x major providers (China Telecom, China Mobile, and China Unicom), while smaller service providers interconnect regions, cities, and services \cite{wang_internet_2024}.

However, in this study, the newly identified hop-point had an IP address associated with Jiangsu Sanai Network Science \& Technology Co. Ltd through AS132883 (Hong Kong). Chinese law compels domestic companies to operate congruent to state interests. This IP association was a possible operational security (OPSEC) failure by UAT-7810.

\textbf{Geographic Association} When reviewing the background of Jiangsu Sanai Network Science \& Technology Co. Ltd., this company is a network service provider and handles internet service, cloud services, and hosting. Sanai is a cultural term typically synonymous with "love of motherland, labor, and science". Additionally, it is within 3 km of the Hangzhou Yangtze Delta Big Data Research Center which is a cross-industry incubation center. Both are subject to Chinese domestic security laws. Jiangsu Sanai's naming convention follows stringent Chinese enterprise names \cite{samr_implementing_2023}. On public satellite imagery, this telecommunications company is located within 5 km north (along G205) from Jiangsu SSD and 18 km west (along G312) from the Eastern (Theater Command) Technical Reconnaissance Base (TRB) of the PLA Cyberspace Force (CSF), also co-located in Nanjing \cite{costello_cyberspace_2025}. Now, Jiangsu SSD is the provincial MSS field office for the entire area with a history of enabling cyber operations \cite{united_states_department_of_justice_chinese_2018}. 

\section{Discussion}
\subsection{Interpretation of Results}
\textbf{Hop-Point 5} Based on the numerous non-traditional ports, malicious artifacts in the open directory, naming conventions, and timing, it is highly likely that Hop-Point 5 is an ORB system staged for cyber operations. Hop-Point 5 is likely operated by native Chinese speakers in China based on historical and current selection of nodes and ASN routes.

\textbf{Hop-Point 6} This hop-point has numerous Windows executables and scripts. Some Python scripts do exist, but the system is heavily skewed towards Windows. Although it is a Chisel Client, staging of AdaptixC2 is an indication that the adversary may construct additional C2 channels between VPS.

\textbf{Hop-Point 7} HP7 contains attack scripts to enumerate a target system. It attempts to pass error logs to another VPS server abroad (Hop-Point 9) and send other data back to Alibaba servers (Hop-Point 10) using a spoofed sender IP. It is possible that error logs are delivered elsewhere for faster triage, rather than hopping multiple nodes towards attacker-controlled networks in China.

\textbf{Analysis} The extreme lengths of tradecraft utilized in the malware payloads and staged configuration files cannot be ignored. Hop-Points 5 through 10 are associated with UAT-7810. During the course of research, this team has observed snapshots of these new hop-points over the course of a month and witnessed these IPs slowly converge toward full operating capability.

\textbf{Non-Standard Ports} It is possible that these non-traditional or non-standard alternative ports were a design decision to avoid port-knocking while operations are underway. 

\textbf{Possible Attribution}  UAT-7810 is assessed by Cisco to be a network service provider, a technical architect, and a maintainer of ORB relays for other APT groups to operate from \cite{an_uat-7810_2026}. Jiangsu Sanai Network Science \& Technology Company has the technical access and workforce capable of supporting customers, even if customers potentially dabble in espionage. As observed in previous cases, China's acceleration in the cyber domain has been enabled by a large workforce of contractors \cite{cary_unmasking_2024, noauthor_intrusion_2026}. UAT-7810 provides services to mission-owners and cyber operators congruent to the nation's vision of military-civil fusion. To avoid national embarrassment from cyber forces, China is likely to divest or transfer responsibility of infrastructure to a civilian contractor for plausible deniability.

\subsection{Implications \& Recommendations}
This case study represents an example of a private company enabling cyber attacks.

Based on this case study, this team recommends pivoting on artifact reuse through HTTP open directories on non-standard ports. Adversaries need the C2 directories to facilitate attacks. The pivots must occur following the immediate disappearance of a known or reported ORB hop-point. Blocking IPs is not enough to thwart attacks because adversaries treat IP addresses as ephemeral.

\section{Future Work}
This team will continue to track the APT Group UAT-7810's LapDogs ORB network and analyze artifacts recovered from this proxy network (Hop-Points 1-10), with the intention of producing a larger journal. Additionally, the team will continue to forward suspicious ORB hop-points to law enforcement congruent to public safety as well as make the UAT-7810 dataset available to the academic community, upon request and proper vetting.

\section{Limitations}
Reporting the IP address may lead to a change of TTPs by the threat actor UAT-7810, but this is an accepted risk for the sake of public awareness and public good. The team acknowledges that Hunt.io is a third-party tool and the team has no control over how open directories are discovered by Hunt.io. There are also observational or cognitive biases because this team is composed of American cyber security analysts during a time of heightened geopolitical tensions.

\section{Conclusion}
ORB networks represent a highly ephemeral architecture and an anti-forensic evolution in a complex cyber landscape. The Confetti Magic case study successfully identified a new ORB hop-point and analyzed its capabilities and behavior. This case study captures the transition between deprecated ORB systems upon discovery by the cyber security community and formation of Hop-Points 5-10. Identification of this hop-point demonstrates evidence potentially linking the new hop-points to the infrastructure provider APT Group UAT-7810. Based on this case study, UAT-7810 is likely a contractor organization for Jiangsu SSD and the Eastern Theater Command Technical Reconnaissance Base of the PLA-CSF. The findings demonstrate that ORB networks consist of sophisticated proxy nodes that can readily rotate and replace one another. By pivoting on behavioral artifact reuse and static naming conventions, defenders can effectively track ORB hop-points in attack campaigns.


\printbibliography

@misc{an_uat-7810_2026,
	title = {{UAT}-7810 continues building {ORB} networks using new malware},
	url = {https://blog.talosintelligence.com/uat-7810/},
	language = {en},
	urldate = {2026-09-09},
	journal = {Cisco Talos Blog},
	author = {An, Jungsoo and Malhotra, Asheer and Svajcer, Vanja and White, Brandon},
	month = jul,
	year = {2026},
}

@techreport{securityscorecard_checking_2025,
	title = {"{Checking} all the {Boxes}" {LapDogs}, {The} {New} {ORB} in {Town}},
	url = {https://securityscorecard.com/wp-content/uploads/2025/06/LapDogs-STRIKE-Report-June-2025.pdf},
    author = {{SecurityScorecard}},
	institution = {SecurityScorecard},
	month = jun,
	year = {2025},
	pages = {26},
}

@article{hwang_exploratory_2023,
	title = {An {Exploratory} {Study} on {Artifacts} for {Cyber} {Attack} {Attribution} {Considering} {False} {Flag}: {Using} {Delphi} and {AHP} {Methods}},
	volume = {11},
	doi = {10.1109/ACCESS.2023.3295427},
	journal = {IEEE Access},
	author = {Hwang, Soonil and Kim, Tae-Sung},
	year = {2023},
	pages = {74533--74544},
}

@article{lee_mucamp_2025,
	title = {{MuCamp}: {Generating} {Cyber} {Campaign} {Variants} via {TTP} {Synonym} {Replacement} for {Group} {Attribution}},
	volume = {20},
	copyright = {https://ieeexplore.ieee.org/Xplorehelp/downloads/license-information/IEEE.html},
	issn = {1556-6013, 1556-6021},
	shorttitle = {{MuCamp}},
	url = {https://ieeexplore.ieee.org/document/11029111/},
	doi = {10.1109/TIFS.2025.3578233},
	language = {en},
	urldate = {2026-09-26},
	journal = {IEEE Transactions on Information Forensics and Security},
	author = {Lee, Insup and Choi, Changhee},
	year = {2025},
	pages = {6162--6174},
}

@article{kim_mucamp2_2026,
	title = {{MuCamp2}: {Generating} {Validated} {Cyber} {Campaign} {Variants} via {Constrained} {LLMs} for {Group} {Attribution}},
	volume = {14},
	doi = {10.1109/ACCESS.2026.3700808},
	journal = {IEEE Access},
	author = {Kim, Keunho and Lee, Jimin and Lee, Insup and Han, Insung and Lee, Jungsik and Shim, Shinwoo and Choi, Changhee},
	year = {2026},
	pages = {87329--87344},
}

@article{ge_threatmamba_2026,
	title = {{ThreatMAMBA}: {Achieving} {High}-{Robustness} {Cyber} {Threat} {Attribution} {During} the {Evolution} of {Attacks}},
	volume = {21},
	doi = {10.1109/TIFS.2026.3685967},
	journal = {IEEE Transactions on Information Forensics and Security},
	author = {Ge, Wenhan and Wang, Junfeng and Cui, Zeyuan and Fang, Zhiyang and Zhan, Weilu},
	year = {2026},
	pages = {4386--4401},
}

@inproceedings{zhang_ephemeral_2020,
	address = {Valencia, Spain},
	title = {Ephemeral {Exit} {Bridges} for {Tor}},
	copyright = {https://ieeexplore.ieee.org/Xplorehelp/downloads/license-information/IEEE.html},
	isbn = {978-1-7281-5809-9},
	url = {https://ieeexplore.ieee.org/document/9153401/},
	doi = {10.1109/DSN48063.2020.00042},
	language = {en},
	urldate = {2026-09-26},
	booktitle = {2020 50th {Annual} {IEEE}/{IFIP} {International} {Conference} on {Dependable} {Systems} and {Networks} ({DSN})},
	publisher = {IEEE},
	author = {Zhang, Zhao and Vaidya, Tavish and Subramanian, Kartik and Zhou, Wenchao and Sherr, Micah},
	month = jun,
	year = {2020},
	pages = {253--265},
}

@article{chen_defending_2025,
	title = {Defending {Against} {APT} {Attacks} in {Cloud} {Computing} {Environments} {Using} {Grouped} {Multiagent} {Deep} {Reinforcement} {Learning}},
	volume = {12},
	doi = {10.1109/JIOT.2025.3542119},
	number = {12},
	journal = {IEEE Internet of Things Journal},
	author = {Chen, Jialin and Lan, Xiaolong and Zhang, Qingquan and Ma, Wengang and Fang, Wenbo and He, Junjiang},
	year = {2025},
	pages = {19459--19470},
}

@inproceedings{zuo_empirical_2025,
	title = {An {Empirical} {Study} on the {Multi}-{Stage} {Nature} of {APT} {Attacks} in {Cloud} {Computing}},
	doi = {10.1109/MASS66014.2025.00095},
	booktitle = {2025 {IEEE} 22nd {International} {Conference} on {Mobile} {Ad}-{Hoc} and {Smart} {Systems} ({MASS})},
	author = {Zuo, Fei and Rhee, Junghwan and Lu, Shuaibing and Song, Yuqi and Zhang, Xin},
	year = {2025},
	pages = {585--591},
}

@inproceedings{poornachandran_internet_2015,
	title = {Internet of {Vulnerable} {Things} ({IoVT}): {Detecting} {Vulnerable} {SOHO} {Routers}},
	doi = {10.1109/ICIT.2015.49},
	booktitle = {2015 {International} {Conference} on {Information} {Technology} ({ICIT})},
	author = {Poornachandran, Prabaharan and Sreeram, R. and Krishnan, Manu R. and Pal, Soumajit and Prem Sankar, A. U. and Ashok, Aravind},
	year = {2015},
	pages = {119--123},
}

@inproceedings{jain_accounting_2023,
	title = {Accounting for the {Unaccounted} {Vulnerabilities} {Found} in {Endpoint} {ISP} {SOHO} {Router}},
	doi = {10.1109/ICCST59048.2023.10474274},
	booktitle = {2023 {IEEE} {International} {Carnahan} {Conference} on {Security} {Technology} ({ICCST})},
	author = {Jain, Akshat and Shete, Prasanna},
	year = {2023},
	pages = {1--4},
}

@article{su_covert_2021,
	title = {Covert {Communication} {With} {Relay} {Selection}},
	volume = {10},
	copyright = {https://ieeexplore.ieee.org/Xplorehelp/downloads/license-information/IEEE.html},
	issn = {2162-2337, 2162-2345},
	url = {https://ieeexplore.ieee.org/document/9241054/},
	doi = {10.1109/LWC.2020.3033786},
	number = {2},
	urldate = {2026-10-20},
	journal = {IEEE Wireless Communications Letters},
	author = {Su, Yinjie and Sun, Hongjian and Zhang, Zhenkai and Lian, Zhuxian and Xie, Zhibin and Wang, Yajun},
	month = feb,
	year = {2021},
	pages = {421--425},
}

@article{zou_passive_2026,
	title = {A {Passive} {Network} {Storage} {Covert} {Channel} for {Internet} of {Things}},
	volume = {13},
	doi = {10.1109/JIOT.2026.3663872},
	number = {9},
	journal = {IEEE Internet of Things Journal},
	author = {Zou, Zexiao and Wang, Zhiqiang and Huang, Qianli and Liu, Baoxu},
	year = {2026},
	pages = {19600--19615},
}

@article{ding_covert_2025,
	title = {Covert {Communication} {With} {Incumbent} {Channel} {Activities}},
	volume = {14},
	doi = {10.1109/LWC.2023.3294343},
	number = {1},
	journal = {IEEE Wireless Communications Letters},
	author = {Ding, Haichuan and An, Jianping},
	year = {2025},
	pages = {3--7},
}

@misc{united_states_cyber__infrastructure_security_agency_defending_2026,
	type = {Cybersecurity {Advisory}},
	title = {Defending {Against} {China}-{Nexus} {Covert} {Networks} of {Compromised} {Devices}},
	url = {https://www.cisa.gov/news-events/cybersecurity-advisories/aa26-113a},
	language = {en},
	urldate = {2026-10-16},
	journal = {United States Cyber \& Infrastructure Security Agency},
	author = {{United States Cyber \& Infrastructure Security Agency}},
	month = apr,
	year = {2026},
}

@misc{united_states_department_of_justice_chinese_2018,
	title = {Chinese {Intelligence} {Officers} and {Their} {Recruited} {Hackers} and {Insiders} {Conspired} to {Steal} {Sensitive} {Commercial} {Aviation} and {Technological} {Data} for {Years}},
	url = {https://www.justice.gov/archives/opa/pr/chinese-intelligence-officers-and-their-recruited-hackers-and-insiders-conspired-steal},
	language = {en},
	urldate = {2026-10-16},
	journal = {Office of Public Affairs},
	author = {{United States Department of Justice}},
	month = oct,
	year = {2018},
}

@techreport{federal_bureau_of_investigation_peoples_2024,
	type = {Joint {Cybersecurity} {Advisory}},
	title = {People’s {Republic} of {China}-{Linked} {Actors} {Compromise} {Routers} and {IoT} {Devices} for {Botnet} {Operations}},
	number = {JCSA-20240918-001},
	institution = {Federal Bureau of Investigation (FBI)},
	author = {{Federal Bureau of Investigation} and {Cyber National Mission Force} and {National Security Agency}},
	month = sep,
	year = {2024},
}

@misc{team_cymru_introduction_2024,
	title = {An {Introduction} to {Operational} {Relay} {Box} ({ORB}) {Networks}},
	url = {https://www.team-cymru.com/post/an-introduction-to-operational-relay-box-orb-networks-unpatched-forgotten-and-obscured},
	language = {en},
	urldate = {2026-10-16},
	journal = {Team Cymru},
	publisher = {Team Cymru},
	author = {{Team Cymru}},
	month = oct,
	year = {2024},
}

@techreport{vanderlee_chinas_2022,
	type = {Threat {Intelligence}},
	title = {China’s {Capabilities} for {State}-{Sponsored} {Cyber} {Espionage}},
	url = {https://www.uscc.gov/sites/default/files/2022-02/Kelli_Vanderlee_Testimony.pdf},
	language = {en},
	institution = {Mandiant},
	author = {Vanderlee, Kelli},
	month = feb,
	year = {2022},
}

@misc{raggi_ioc_2024,
	title = {{IOC} {Extinction}? {China}-{Nexus} {Cyber} {Espionage} {Actors} {Use} {ORB} {Networks} to {Raise} {Cost} on {Defenders}},
	shorttitle = {{IOC} {Extinction}?},
	url = {https://cloud.google.com/blog/topics/threat-intelligence/china-nexus-espionage-orb-networks},
	language = {en},
	urldate = {2026-10-16},
	journal = {Google Cloud Blog},
	publisher = {Google},
	author = {Raggi, Michael},
	month = may,
	year = {2024},
}

@misc{mandiant_intelligence_stealth_2023,
	title = {Stealth {Mode}: {Chinese} {Cyber} {Espionage} {Actors} {Continue} to {Evolve} {Tactics} to {Avoid} {Detection}},
	shorttitle = {Stealth {Mode}},
	url = {https://cloud.google.com/blog/topics/threat-intelligence/chinese-espionage-tactics},
	language = {en},
	urldate = {2026-10-16},
	journal = {Google Cloud Blog},
	publisher = {Google},
	author = {{Mandiant Intelligence}},
	month = jul,
	year = {2023},
}

@misc{bianco_enterprise_2013,
	title = {Enterprise {Detection} \& {Response}: {The} {Pyramid} of {Pain}},
	shorttitle = {Enterprise {Detection} \& {Response}},
	url = {https://detect-respond.blogspot.com/2013/03/the-pyramid-of-pain.html},
	urldate = {2026-10-16},
	journal = {Enterprise Detection \& Response},
	author = {Bianco, David J.},
	month = mar,
	year = {2013},
}

@techreport{brezinski_guidelines_2002,
	type = {Request for {Comments}},
	title = {Guidelines for {Evidence} {Collection} and {Archiving}},
	url = {https://datatracker.ietf.org/doc/rfc3227},
	doi = {10.17487/RFC3227},
	number = {RFC 3227},
	urldate = {2026-10-16},
	institution = {Internet Engineering Task Force},
	author = {Brezinski, D. and Killalea, T.},
	month = feb,
	year = {2002},
	note = {Num Pages: 10},
}

@inproceedings{10.1145/3689932.3694766,
author = {Shashwat, Kumar and Hahn, Francis and Millar, Stuart and Ou, Xinming},
title = {Using LLM Embeddings with Similarity Search for Botnet TLS Certificate Detection},
year = {2024},
isbn = {9798400712289},
publisher = {Association for Computing Machinery},
address = {New York, NY, USA},
url = {https://doi.org/10.1145/3689932.3694766},
doi = {10.1145/3689932.3694766},
booktitle = {Proceedings of the 2024 Workshop on Artificial Intelligence and Security},
pages = {173–183},
numpages = {11},
location = {Salt Lake City, UT, USA},
series = {AISec '24}
}

@inproceedings{10.1145/3727967.3756837,
author = {Alomari, Dorieh and Alabdullatif, Maryam Ahmed and Khan, Fakhri Alam},
title = {An Explainable AI-based Network Intrusion Detection System for Botnet Attacks},
year = {2025},
isbn = {9798400718328},
publisher = {Association for Computing Machinery},
address = {New York, NY, USA},
url = {https://doi.org/10.1145/3727967.3756837},
doi = {10.1145/3727967.3756837},
booktitle = {Proceedings of the 2025 29th International Conference on Evaluation and Assessment in Software Engineering Companion},
pages = {169–175},
numpages = {7},
location = {Istanbul, Turkiye},
series = {EASE Companion '25}
}

@techreport{healey_beyond_2012,
	type = {Issue brief},
	title = {Beyond {Attribution}: {Seeking} {National} {Responsibility} for {Cyber} {Attacks}},
	url = {https://www.atlanticcouncil.org/wp-content/uploads/2012/02/022212_ACUS_NatlResponsibilityCyber.PDF},
	language = {en},
	institution = {Atlantic Council},
	author = {Healey, Jason},
	year = {2012},
}

@book{xiao_science_2022,
	address = {Beijing},
	title = {The {Science} of {Military} {Strategy} (2020)},
	isbn = {979-8-4050-7762-8},
	publisher = {China Aerospace Studies Institute},
	editor = {Xiao, Tianliang},
	translator = {{China Aerospace Studies Institute}},
	year = {2022},
	note = {original-date: 2020},
}

@misc{noauthor_matter_2024,
	title = {In the {Matter} of the {Search} of {Infected} {Devices} and {Command}-and-{Control} {Servers} in the {United} {States} {That} {Are} {Part} of a {Mirai} {Botnet}},
	url = {https://www.justice.gov/archives/opa/media/1369156/dl/dl?inline},
	month = sep,
	year = {2024},
}

@techreport{cna_chinas_2023,
	address = {Arlington, VA},
	title = {China's {National} {Security} {Laws}: {Implications} {Beyond} {Borders}},
	url = {https://www.cna.org/quick-looks/2023/China-national-security-laws-implications-beyond-borders.pdf},
	number = {DMM-2023-U-037195-Final},
	institution = {CNA},
	author = {{CNA}},
	month = dec,
	year = {2023},
}

@misc{ying_as_2025,
	address = {Singapore},
	title = {As the {West} {Turns} {Cold}, {China}’s {Tech} {Giants} {Warm} {Up} to {Southeast} {Asia} - {With} {Deep} {Pockets}},
	url = {https://www.channelnewsasia.com/east-asia/china-tech-giants-tencent-alibaba-sensetime-pivot-southeast-asia-5280276},
	publisher = {CNA},
	author = {Ying, Bong Xin},
	month = nov,
	year = {2025},
}

@misc{wang_internet_2024,
	address = {Wellington},
	type = {Conference {Presentation}},
	title = {Internet {Interconnection} {Architecture} of {China}},
	url = {https://apix.asia/30/apix30-wang-architecturechina.pdf},
	author = {Wang, Qifeng},
	month = sep,
	year = {2024},
}

@misc{samr_implementing_2023,
	title = {Implementing {Measures} for the {Provisions} on the {Administration} of {Enterprise} {Name} {Registration}},
	url = {https://www.samr.gov.cn/zw/zfxxgk/fdzdgknr/fgs/art/2023/art_1e269e76abdb405ab5253b7c78e45f6a.html},
	language = {zh},
    author = {{SAMR}},
	month = aug,
	year = {2023},
}

@misc{thomas_comey_2014,
	title = {Comey: {Chinese} hackers like a 'drunk burglar'},
	shorttitle = {Comey},
	url = {https://apnews.com/united-states-government-af0de19bb2784b6093658a1f5770a4f0},
	language = {en},
	urldate = {2026-10-20},
	journal = {AP News},
	author = {Thomas, Ken},
	month = oct,
	year = {2014},
}

@misc{noauthor_intrusion_2026,
	title = {Intrusion {Truth}},
	url = {https://intrusiontruth.wordpress.com/},
	urldate = {2026-10-22},
	year = {2026},
}

@misc{cary_unmasking_2024,
	title = {Unmasking {I}-{Soon} {\textbar} {The} {Leak} {That} {Revealed} {China}’s {Cyber} {Operations}},
	url = {https://www.sentinelone.com/labs/unmasking-i-soon-the-leak-that-revealed-chinas-cyber-operations/},
	language = {en},
	urldate = {2026-10-22},
	journal = {SentinelOne},
	author = {Cary, Dakota and Milenkoski, Aleksandar},
	month = feb,
	year = {2024},
}

@misc{costello_cyberspace_2025,
  title = {The {Cyberspace} {Force}: {A} {Bellwether} for {Conflict}},
  shorttitle = {The {Cyberspace} {Force}},
  url = {https://jamestown.org/the-cyberspace-force-a-bellwether-for-conflict/},
  language = {en},
  urldate = {2026-10-22},
  journal = {Jamestown},
  author = {Costello, John},
  month = apr,
  year = {2025},
}

\end{document}